\documentclass[journal,transmag]{IEEEtran}
\usepackage{mathtools}
\usepackage{float}
\usepackage{graphicx} 
\usepackage[colorinlistoftodos]{todonotes}
\graphicspath{ {images/} }

\usepackage[linesnumbered,ruled,vlined]{algorithm2e} 

\usepackage[utf8]{inputenc}
\usepackage[english]{babel}
\usepackage{amsthm,amssymb}

\usepackage{optidef}

\theoremstyle{definition}

\IEEEoverridecommandlockouts                              

\title{
List Viterbi Decoding of PAC Codes
}

\author{
\IEEEauthorblockN{Mohammad Rowshan, {\em Student Member, IEEE} and Emanuele Viterbo, {\em Fellow, IEEE}}
 \thanks{M. Rowshan and E. Viterbo are with the  Department of Electrical and Computer Systems Engineering (ECSE), Monash University, Melbourne, VIC3800, Australia. E-mail: \{mohammad.rowshan, emanuele.viterbo\}@monash.edu. These authors' work was supported by the Australian Research Council under Discovery Project ARC DP160100528.}
}

\begin{document}

\maketitle
\thispagestyle{empty}
\pagestyle{empty}

\begin{abstract}
Polarization-adjusted convolutional (PAC) codes are special concatenated codes in which we employ a one-to-one convolutional transform as a pre-coding step before the polar transform. In this scheme, the polar transform (as a mapper) and the successive cancellation process (as a demapper) present a synthetic vector channel to the convolutional transformation. The numerical results show that this concatenation improves the Hamming distance properties of polar codes. 
In this work, we implement the parallel list Viterbi algorithm (LVA) and show how the error correction performance moves from the poor performance of the Viterbi algorithm (VA) to the superior performance of list decoding by changing the constraint length, list size, and the sorting strategy (local sorting and global sorting) in the LVA. Also, we analyze the latency of the local sorting of the paths in LVA relative to the global sorting in the list decoding and the trade-off between the sorting latency and the error correction performance.
\end{abstract}


\section{Introduction}
\label{sec:intro}
Polar codes proposed by Ar\i kan in \cite{arikan} are the first class of  channel codes with an  explicit construction that was proven to achieve the symmetric (Shannon) capacity of a binary-input discrete memoryless channel (BI-DMC) using a low-complexity successive cancellation (SC) decoder (SCD). Nevertheless, 
the error correction performance of finite-length polar codes under SCD is not satisfactory due to the existence of partially polarized channels. To address this issue, SC list decoding (SCLD or in short LD) was proposed in \cite{tal}. 

\begin{figure}
    \centering
    \includegraphics[width=1\columnwidth]{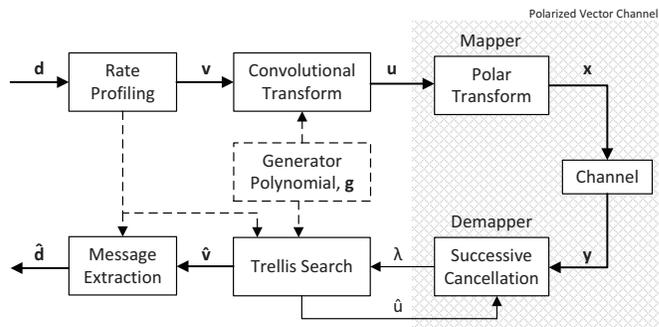}
    \caption{\vspace{-5mm} PAC Coding Scheme} 
    \label{fig:PAC_scheme}
\end{figure}

Recently in \cite{arikan2}, Ar\i kan proposed a concatenation of a convolutional transform with the polarization transform \cite{arikan} in which a message is first encoded using a convolutional transform and then transmitted over polarized synthetic channels as shown in Fig.~\ref{fig:PAC_scheme}. These codes are called ``polarization-adjusted convolutional (PAC) codes". 
It was shown in \cite{li2} that a properly designed pre-transformation such as convolutional transform can improve the distance properties of polar codes.

In \cite{rowshan-pac1}, we studied the  implementation of tree search algorithms including the conventional list decoding and complexity-efficient Fano decoding for PAC codes. However, due to the convolutional pre-transformation, PAC codes can also be easily encoded and decoded based on the trellis by employing the Viterbi algorithm (VA) \cite{viterbi,forney} and the extended/list VA \cite{hashimoto} as a decoder. The basic Viterbi algorithm was employed in \cite{arikan_viterbi} as an ML decoder for short polar codes in a comparison with Reed-Muller (RM) codes. 

In this paper, we illustrate the implementation of the parallel list Viterbi algorithm (LVA) \cite{hashimoto,seshadri} for PAC codes, and analyse the impact of list size and constraint length on the error correction performance. We also analyze the latency of the path sorting at each state on the trellis relative to global sorting in tree-based list decoding.


\section{Preliminaries}\label{sec:prelim}

Polarization-adjusted convolutional (PAC) codes are denoted by {\em PAC}$(N,K,\mathcal{A},\mathbf{g})$, where $N = 2^n$ is the length of the PAC code. A rate profiler first maps the $K$ information bits to $N$ bits. Then, the convolutional transform (with polynomial coefficients vector $\mathbf{g}$) scrambles the resulting $N$ bits before feeding them to the classical polar transform (Fig.~\ref{fig:PAC_scheme}). 
The information bits $\mathbf{d}=[d_0,d_1,...,d_{K-1}]$ are interspersed with $N-K$ zeros and mapped to the vector $\mathbf{v}=[v_0,v_1,...,v_{N-1}]$ using a rate-profile which defines the code construction. The rate-profile is defined by the index set $\mathcal{A}\subseteq \{0,\ldots, N-1\}$, where the information bits appear in $\mathbf{v}$. This set can be defined as the indices of sub-channels in the polarized vector channel with high reliability. 
These sub-channels are called {\em good channels}. The bit values in the remaining positions $\mathcal{A}^c$ in $\mathbf{v}$ are set to 0.  

The input vector $\mathbf{v}$ is transformed to vector $\mathbf{u}=[u_0,\ldots,u_{N-1}]$ as $u_i = \sum_{j=0}^m g_j v_{i-j}$ using the binary generator polynomial of degree $m$, with coefficients $\mathbf{g}=[g_0,\ldots,g_m]$. 
The convolutional transform combines $m$ previous input bits stored in a shift register with the current input bit $v_i$ to calculate $u_i$ (see subroutine {\em conv1bEnc} in Algorithm \ref{alg:LVA}). The parameter $m+1$, in bits, is called the {\em constraint length} of the convolutional code. As a result of this pre-transformation, $u_i$ for $i\in\mathcal{A}^c$ are no longer frozen as in polar codes.
Note that this convolutional transformation is one-to-one, therefore the output vector $\mathbf{u}$ is not a traditional convolutional codeword. The rate-profiling process performed before the convolutional transformation creates the redundancy by inserting $N-K$ zeros in the length-$K$ input sequence $\mathbf{d}$.  

Finally, as shown in Fig.~\ref{fig:PAC_scheme}, vector $\mathbf{u}$ is mapped to vector $\mathbf{x}$ ($\mathbf{x}=\mathbf{u}\mathbf{P}_n$) by the polar transform $\mathbf{P}_n=\mathbf{P}^{\otimes n}$ defined as the $n$-th Kronecker power of  
$\mathbf{P} = 
{\footnotesize \begin{bmatrix}
1 & 0 \\
1 & 1
\end{bmatrix} }$. 

The $\mathbf{x}$ vector is transmitted through a noisy channel and received as the vector $\mathbf{y}$. The channel log-likelihood ratios (channel LLRs) computed based on the received signals $\mathbf{y}$ by $\lambda_n^t= \ln\frac{P(Y_t=y_i|X_t=+1)}{P(Y_i=y_t|X_t=-1)}=\frac{2}{\sigma^2}y_t$. The outputs of demapping by successive cancellation process are denoted by $\lambda_0^{0,N-1}$ which are simply shown by $\lambda$ in Fig. \ref{fig:PAC_scheme}. Note that the subscript $n$ and $0$ in $\lambda_0^{n,N-1}$ and $\lambda_0^{0,N-1}$ denote respectively the first and the last stages of the SC factor graph  shown in Fig. 1 of \cite{rowshan-stepped}.  In the next section, we describe the decoding process and define the path metric.

\section{List Viterbi Decoding}\label{sec:LVA}
The Viterbi algorithm \cite{viterbi} is the most popular decoding procedure for convolutional codes (CCs), which is based on their {\em trellis diagram}  graphical representation \cite{forney}. A trellis is a directed graph where the nodes represent the encoder state. The branch sequences on the trellis are generated by a finite state machine with inputs $\mathbf{v}$ and states $\mathcal{S}=\{s_1,...,s_{2^m}\}$ and the code is called the trellis code. The Viterbi algorithm traverses the trellis from left to right, finding the maximum likelihood transmitted sequence estimate, when reaching the last stage $t=N-1$.

\begin{figure}
    \centering
    \includegraphics[width=1\columnwidth]{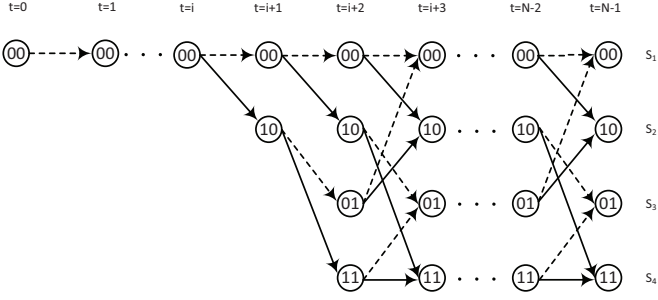}
    \caption{The truncated trellis for PAC codes. Since $v_t=0$ for $t\in \mathcal{A}^c$, 
    the path does not split. The dashed-line arrows represent the input 0 and the solid-line arrows represent the input 1 to the convolutional transform.} 
    \label{fig:irregular_trellis1}
\end{figure}
\begin{figure}
    \centering
    \includegraphics[width=1\columnwidth]{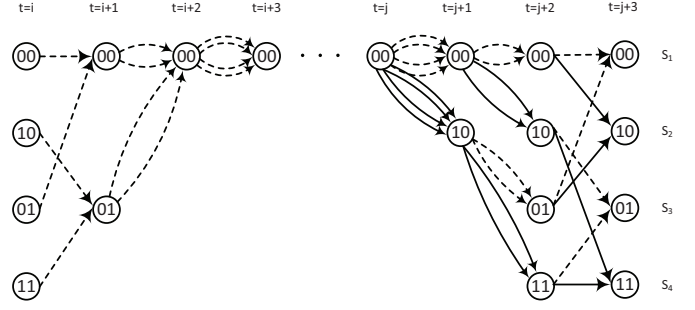}
    \caption{The irregularity of the trellis where $v_t=0$ for $t\in \mathcal{A}^c$ or $t=[i+1,...,j]$ for $j>i$. The paths from $t=i+1$ to $t=j$ are not pruned.
    } 
    \label{fig:irregular_trellis2}
\end{figure}

PAC codes can be encoded and decoded on the trellis. The trellis used for PAC codes is an irregular trellis which is shown in Fig. \ref{fig:irregular_trellis1} and  \ref{fig:irregular_trellis2}. As shown, when there is a sub-sequence of at least $m$ zeros in the input $\mathbf{v}$, the current states of all the paths on the trellis transit toward all-zero state. 

In convolutional coding, there are three methods to obtain the finite code sequences: (1) code truncation where the encoder stops  after a finite block-length, $N$, and the code sequence is truncated. This method leads to a substantial degradation of error protection, because the last encoded information bits influence a small number of code bits. (2) code termination where we add some tail bits to the code sequence in order to ensure a predefined end state (usually, the all-zero state) of the encoder, which leads to low error probabilities for the last bits, (3) tail-biting where we choose a starting state that ensures the starting and ending states are the same (this state value does not necessarily have to be the all-zero state). This scheme avoids the rate loss incurred by zero-tail termination at the expense of a more complex decoder. 
For encoding PAC codes, we use the code truncation, thus we do not add any tail bits. This will not degrade the error protection of last bits because the last encoded bits are transmitted over the high-reliability sub-channels in the polar transform. 

The fundamental idea behind the Viterbi decoding is as follows. A coded sequence $\mathbf{u}$, the output of the convolutional transform in Fig. \ref{fig:PAC_scheme}, corresponds to a path through the trellis. Due to the noise in the channel, the received vector $\mathbf{y}$ after demapping may not correspond exactly to a path on the trellis. The decoder finds a path through the trellis which has the highest probability to be the transmitted sequence $\mathbf{u}$ over the polarized vector channel. 
The probability to be {\em{maximized}} is

\begin{equation} 
\label{eq:pac_metric1}
\begin{multlined}
P(\hat{\mathbf{u}}|\mathbf{y})=\prod_{t=0}^{N-1} P(\hat{u}_t|\hat{u}_0^{t-1},y_0^{N-1})
\end{multlined}
\end{equation}

In practice, it is convenient to deal with the logarithm of (\ref{eq:pac_metric1}) to use an additive metric. Consider now a partial sequence   $\hat{u}_0^{t-1}=[\hat{u}_0, \hat{u}_1,\ldots, \hat{u}_{t-1}]$ at the output of the convolutional transform. This sequence determines a path, or a sequence of states, through the trellis for the code. 

Let $M_{t-1}(s^\prime)=-\sum_{i=0}^{t-1}\log P(\hat{u_i}|\hat{u}_0^{i-1},y_0^{N-1})$ denote the {\em path metric} for the sequence $\hat{u}_0^{t-1}$ terminating in state $s^\prime$. 
We seek to minimize the path metric for the entire codewords ($t=N-1$) to maximize the probability in (\ref{eq:pac_metric1}). 

Now let the sequence $\hat{u}_0^{t}$ be obtained by appending  $\hat{u}_t$ to $\hat{u}_0^{t-1}$ and suppose $\hat{u}_t$ is such that the state at time $t + 1$ is $s$. The path metric for this longer sequence is

\begin{eqnarray} 
M_t(s)&=&-\sum_{i=0}^{t}\log P(\hat{u}_i|\hat{u}_0^{i-1},y_0^{N-1})
\label{eq:pac_metric2}\\
&=&M_{t-1}(s^\prime)+\mu_t(s^\prime,s) \label{eq:pac_metric3}
\end{eqnarray}
where $\mu_t(s^\prime,s)=-\log P(\hat{u}_t|\hat{u}_0^{t-1},y_0^{N-1})$ denotes the {\em branch metric} for the trellis transition from state $s'$ at time $t$ to state $s$ at time $t + 1$.


The path metric along a path to state $s$ at time $t$ is obtained by adding the path metric
to the state $s^\prime$ at time $t - 1$ to the branch metric for an input that moves the encoder from
state $s^\prime$ to state $s$. If there is no such input, i.e., $s^\prime$ and $s$ are not connected on the trellis, then the branch metric is considered $\infty$.

To simplify the arithmetic operation, we can define $\mu_t$ based on $\lambda_0^t(s^\prime,s)$ or simply $\lambda_0^t$. 
\begin{equation}
\label{eq:pac_metric4}
\begin{multlined}
\mu_t(s^\prime,s) = -\log P(\hat{u}_t|\hat{u}_0^{t-1},y_0^{N-1})\\=-\log\left(\frac{e^{(1-\hat{u}_t)\lambda_0^t}}{e^{\lambda_0^t}+1}\right) 
=\log\left(1+e^{-(1-2\hat{u}_t)\lambda_0^t}\right)
\end{multlined}
\end{equation}
where the last equality holds only for $\hat{u}_t=\hat{u}_t(s^\prime,s)=$ 0 and\,1. 
Now, for the value of $\hat{u}_t$ that equals  $h(\lambda_0^t)$,
\begin{equation}
\label{eq:sc_hard_decision}
h(\lambda_0^t) = \begin{dcases*}
        0 & $\lambda_0^t>0$,\\
        1 & otherwise\\
\end{dcases*}
\end{equation}
the term $e^{-(1-2\hat{u}_t)\lambda_0^t}=e^{-|\lambda_0^t|}$ is small and hence $\log(1+~e^{-|\lambda_0^t|}) \approx 0$. Otherwise, we can approximate $\log(1+e^{|\lambda_0^t|})\approx |\lambda_0^t|$. Thus

\begin{equation}
\label{eq:pm_func} 
\mu_t(s^\prime,s)=\mu_t(\lambda_0^t,\hat{u}_t)\! \approx \!
\begin{dcases*}
0 & if $\hat{u}_t = h(\lambda_0^t)$\\ 
|\lambda^t_0| & otherwise	\\
\end{dcases*}
\end{equation}

It turns out that this branch metric is equivalent to the one suggested for the list decoding of polar codes in \cite{yuan,balatsoukas} and PAC codes in \cite{rowshan-pac1}.

\begin{figure}[t]
    \centering
    \includegraphics[width=0.8\columnwidth]{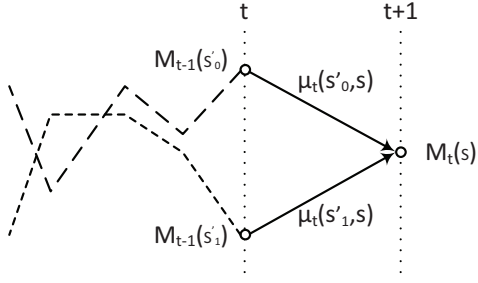}
    \caption{\vspace{-5mm} Merging two paths at state $s$} 
    \label{fig:paths_merge}
\end{figure}

When paths merge at state $s$, we need to select one of them in order to  extend it at the next time step. Suppose $M_{t-1}(s^\prime_0)$ and $M_{t-1}(s^\prime_1)$ are the path metrics of the paths ending at states $s^\prime_0,s^\prime_1\in \{0,1,...,2^m-1\}$ at time $t$. Suppose further that both of these states are connected to state $s$ at time $t + 1$, as illustrated in Fig. \ref{fig:paths_merge}. 

According to the Bellman's principle of optimality \cite{belman}, to obtain the maximum likelihood (ML) path through the trellis, the path to any state at each time step must be locally an ML path. 
This is the governing principle of the Viterbi algorithm. Thus, when the two or more paths merge, the path with the smallest path metric is retained (the {\em survivor path} or in short the {\em survivor}) and the other path is eliminated from further consideration. 
This defines the {\em add-compare-select} step of the Viterbi algorithm
\begin{equation}
\label{eq:pac_metric5}
\begin{multlined}
M_t(s) = \min\{ M_{t-1}(s^\prime_0) + \mu_t(s^\prime_0,s),\\ M_{t-1}(s^\prime_1) + \mu_t(s^\prime_1,s) \}
\end{multlined}
\end{equation}

Note that the initial path metrics are $M_0(0) = 0$ and $M_0(s^\prime) = \infty$ for $s^\prime = 1,2, . . . , 2^m - 1$. 

In \cite{hashimoto}, the conventional Viterbi algorithm was generalized to list-type VA where instead of one path, the $L$ paths with smallest metric are selected and extended at time $t$. Hence, (\ref{eq:pac_metric5}) is generalized as 
\begin{equation}
\label{eq:pac_metric_lva}
\begin{multlined}
M_t(s,k) = \min_{\substack{1\leq l\leq L\\s^\prime}}^{(k)}\{ M_{t-1}(s^\prime,l) + \mu_t(s^\prime,s) \}
\end{multlined}
\end{equation}
where $\min^{(k)}$ denotes the $k$-th smallest value ($1\leq k\leq L$). 

From (\ref{eq:pac_metric_lva}), one can observe some similarity between list decoding of PAC codes and list Viterbi algorithm (LVA). The main difference is that in the LVA, the paths are sorted locally at each state, while in list decoding all the paths are sorted globally and then half of them are discarded. 


Algorithm \ref{alg:LVA} illustrates the list Viterbi algorithm. In the beginning, there is a single path in the list. When the index of the current bit is in the set $\mathcal{A}^c$, the decoder knows its value, usually $v_t=0$, and therefore it is encoded into $u_t$ based on the current memory state $S$ and the generator polynomial $\mathbf{g}$ in line 8. 
Then, using the decision LLR $\lambda_0^t$ obtained in line 6, the corresponding path metric is calculated using subroutine $calcM$. Note that in the algorithm \ref{alg:LVA}, instead of $M_t(s,k)$  in \ref{eq:pac_metric_lva}, we use $M_t(k)$.  Although the metric is calculated in lines 9 and 26-27 regardless of the current state of each corresponding path, when we sort the paths in line 16, we consider their current states. Eventually, the decoded value $u_t$ is fed back into SC process in line 10 to calculate  the partial sums. 
On the other hand, if the index of the current bit is in the set $\mathcal{A}$ (see lines 12-17), there are two options for the value of $v_t$, i.e., 0 and 1, to be considered in line 23. For each option of 0 and 1, the aforementioned process for $t\in \mathcal{A}^c$ including convolutional encoding, and calculating the path metric is performed and then the two encoded values $u_t=0$ and $1$ are fed back into SC\,process to update the partial sums $\beta_\pi$. 



The vector $\mathbf{\lambda}[\pi]$ as the input argument of the subroutine {\em updateLLRs}  constitutes the $N-1$ intermediate LLR values of path $\pi$. The subroutine updateLLRs updates all the intermediate LLRs and gives $\lambda_0^t[\pi]$. Similarly, the vector  $\beta_\pi$ constitutes the $N-1$ intermediate partial sums of path $\pi$ which is needed to compute the intermediate LLRs. The partial sums are updated after decoding each bit by the subroutine {\em updatePartialSums}. The subroutines {\em updateLLRs}, {\em updatePartialSums}, and {\em prunePaths} in Algorithm \ref{alg:LVA} are identical to the ones used in SCL decoding of polar codes.

\begin{algorithm}
\label{alg:LVA}
\caption{List Viterbi Decoding of PAC codes}
\DontPrintSemicolon
\SetKwInOut{Input}{input}
\SetKwInOut{Output}{output}
\SetKwRepeat{Repeat}{do}{while} 
\SetKwFunction{func}{Subroutine}
\Input{$\mathcal{A}$, $L$, $\mathbf{g}$, $\lambda_n^{0,N-1}$}
\Output{the recovered message bits $\mathbf{\hat{d}}$}
    $\Pi \gets \{1\}$ \tcp*{\footnotesize a single path in the list}
    $m \gets |\mathbf{g}|-1$ \tcp*{\footnotesize memory size}
    \For{$t\gets 0$ \KwTo $N-1$}{
        \uIf{$t \notin \mathcal{A}$}{
            \For{$\pi\gets 1$ \KwTo $|\Pi|$}{
                $\lambda_0^t[\pi]\gets $ updateLLRs($\pi$, $t$, $\mathbf{\lambda}[\pi]$, $\mathbf{\beta}_\pi$)\tcp*{\footnotesize updateLLRs: Identical with SCD's}
                $\hat{v}_t[\pi]\gets 0$\;
                {\small [$\hat{u}_t[\pi]$, S[$\pi$]]$\gets $ conv1bEnc(0, S[$\pi$], $\mathbf{g}$)\;}
                $M_t(\pi)\!\gets\!M_{t-1}(\pi)\!+\!\mu_t(\lambda_0^t[\pi], \hat{u}_t[\pi])$  {\footnotesize// cf. \ref{eq:pm_func}}\;
                $\mathbf{\beta}_\pi \gets$ updatePartialSums($\hat{u}_t[\pi]$, $\mathbf{\beta}_\pi$)\tcp*{\footnotesize Identical w/ SCD's}
                
            }
            
        }
        \Else{
            \For{$\pi\gets 1$ \KwTo $|\Pi|$}{
                duplicatePath($\pi$, $t$, $\mathbf{g}$)\;
            }
            \uIf{$|\Pi| > 2^m.L$}{
                \For{$s\gets 1$ \KwTo $2^m$}{
                    Sort \{$M_t(\pi)\}, \pi\in\Pi$ : connected to $s$\;
                    Retain $L$ paths ($\pi$'s) with smallest $M_t$
                }
            }
        }
    }
    $\hat{v}_0^{N-1}[1:|\Pi|] \gets$ sort($\hat{v}_0^{N-1}[1:|\Pi|]$)   // {\footnotesize in ascending order}\;
    $\mathbf{\hat{d}} \gets$ extractData($\hat{v}_0^{N-1}[0]$, $\mathcal{A}$)   // {\footnotesize inverse of  rate-profiling}\;
    \KwRet $\mathbf{\hat{d}}$;

    \SetKwFunction{FdL}{duplicatePath}
    \SetKwProg{Fn}{subroutine}{:}{}
    \Fn{\FdL{$\pi$, $t$, $\mathbf{g}$}}{
        $\Pi\gets\Pi\cup \{\pi^\prime\}$ \tcp*{{\footnotesize path $\pi^\prime$ is a copy of path $\pi$}}
        $\lambda_0^t[\pi]\gets$ updateLLRs($\pi$, $t$, $\mathbf{\lambda}[\pi]$, $\mathbf{\beta}_\pi$)\tcp*{\footnotesize like SCD}
        ($\hat{v}_t[\pi]$, $\hat{v}_t[\pi^\prime]$) $\gets$ (0, 1)\;
        [$\hat{u}_t[\pi]$, S[$\pi$]] $\gets$ conv1bEnc($\hat{v}_t[\pi]$, S[$\pi$], $\mathbf{g}$)\;
        [$\hat{u}_t[\pi^\prime]$, S[$\pi^\prime$]]$\gets$  conv1bEnc($\hat{v}_t[\pi^\prime]$, S[$\pi$], $\mathbf{g}$)\;
        $M_t(\pi)\gets M_{t-1}(\pi) + \mu_t(\lambda_0^t[\pi], \hat{u}_t[\pi])$  \tcp*{\footnotesize cf. \ref{eq:pm_func}}
        $M_t(\pi^\prime)\gets M_{t-1}(\pi) + \mu_t(\lambda_0^t[\pi], \hat{u}_t[\pi^\prime])$  \tcp*{\footnotesize cf. \ref{eq:pm_func}}
        $\mathbf{\beta}_\pi \gets$ updatePartialSums($\hat{u}_t[\pi]$, $\mathbf{\beta}_\pi$)\tcp*{\footnotesize like SCD}
        $\mathbf{\beta}_{\pi^\prime} \gets$ updatePartialSums($\hat{u}_t[\pi^\prime]$, $\mathbf{\beta}_{\pi}$)\;
        
    }
    
    \SetKwFunction{Fceb}{conv1bEnc}
    \SetKwProg{Fn}{subroutine}{:}{}
    \Fn{\Fceb{$v$, {\em currState}, $\mathbf{g}$}}{
        $u \gets v \cdot g_0$\;
        \For{$j\gets 1$ \KwTo $|\mathbf{g}|$}{
            \uIf{$g_j = 1$}{
                $u \gets u$ $\oplus$ currState[$j-1$]\;
            }
        }
        nextState $\gets$ [$v_i$] + currState[1,...,$|\mathbf{g}|-2$]\;
        \KwRet ($u$, nextState);
    }

\end{algorithm}

\section{Generalization of List Viterbi Algorithm}\label{sec:universality}
Successive Cancellation List Viterbi algorithm (SC-LVA or in short LVA) for decoding of PAC codes can be considered a generalized decoder for PAC codes in a sense that it can be converted to SC decoding, SC list decoding and Viterbi decoding by changing the parameters of the algorithm. 

In terms of sorting strategy for the path metrics at each time step, there are two strategies to consider: 
\begin{itemize}
\item 
{\em global sorting} of all the paths regardless of their current states. In this case, LVA will not have a fixed number of survivors for each state (or at each node on the trellis) and the decoding reduces to SC list decoding (LD) of PAC codes. 
In this case, the performance improves by increasing the list size, $L$. A special case of list decoding is SC decoding when $L=1$. 
\item
{\em local sorting} of the paths with the same current state (the paths connected to the same node on the trellis). This is the conventional LVA for PAC codes described in section \ref{sec:LVA}. In this case, by increasing either the list size $L$ or the number of states $|\mathcal{S}|$, while keeping the other parameter constant, the performance improves. However, if we keep the product of $L\cdot|\mathcal{S}|$ constant, an increase in $L$ improves the performance. Note that in this case, if $|\mathcal{S}|$ becomes two small such as $|\mathcal{S}|=2$, the convolution has a limited span and results in a degradation in FER performance as we will see in Section \ref{sec:results}. Needless to mention that if we increase $L\cdot|\mathcal{S}|$, the performance improves.  We note that the PAC code changes by changing $|\mathcal{S}|$, since we are using a different $\mathbf{g}$. Since it was observed that the FER performance of PAc codes is not significantly affected by the change of $\mathbf{g}$, we can vary this parameter and the local list size and observe the tradeoffs of the different decoders.
\end{itemize}
Additionally, when we choose only one path at each state ($L=1$), LVA is converted to a standard Viterbi algorithm (VA) for PAC codes, which was described in Section \ref{sec:LVA}. In this case, as the number of states, $|\mathcal{S}|$, on the trellis increases, the performance improves. Also note that PAC coding with $\mathbf{g}=[1]$ or $m=0$ is equivalent to polar coding simply because there is no pre-transformation or pre-coding in this case.



\section{Sorting Latency} \label{sec:sorting}
As discussed in the previous section, the error correction performance of the decoding changes with the sorting strategy as well as the list size and the number of states. Now, let us analyse the sorting complexity in list decoding and list Viterbi decoding. Suppose the total number of survivor paths is the same in LD and LVA, i.e., $L_{LD}=L_{LVA}\cdot 2^m$. As we will observe in the next section, in the condition of the same number of survivors, LD slightly outperforms LVA due to the global sorting strategy. However, in case of parallelism which is popular in the hardware design, the local sorting in LVA can improve the latency significantly. 

\begin{figure}[t] 
    \centering
    \includegraphics[width=1\columnwidth]{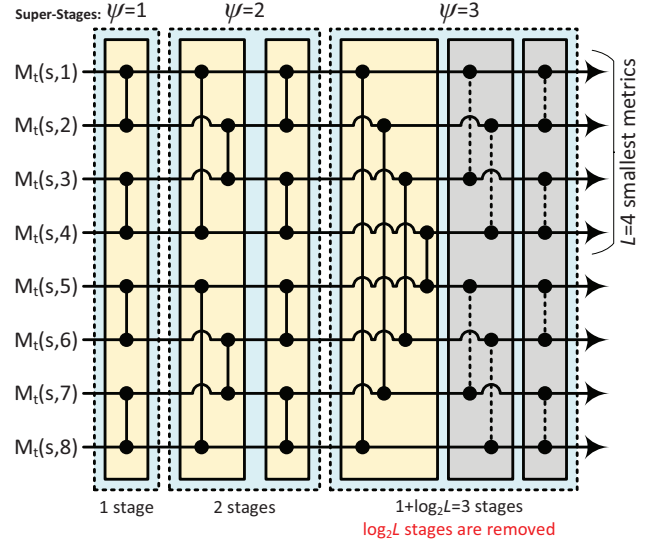}
    \caption{The reduced bitonic sorting network for LVA with $L=4$. The order of $L$ smallest path metrics is not needed.} 
    \label{fig:sorting_network}
\end{figure}

Let us consider a bitonic sorter \cite{batcher} with $1+\log L$ super-stages that can sort $2L$ path metrics shown in Fig. \ref{fig:sorting_network}. At each super-stage with index $\psi\in\{1,...,1+\log L\}$, there are $\psi$ stages (i.e., the number of stages at each super-stage equals the index ($\psi$) of  the corresponding super-stage, see the top and the bottom of Fig. \ref{fig:sorting_network}), each including  $L$ pairs of a component (shown by vertical connections in Fig. \ref{fig:sorting_network})  consists of a comparator and 2-to-2 multiplexer, which work in parallel. This sorter was used for list decoding of polar codes in  \cite{lin} and later improved in \cite{balatsoukas-sorter}.  The length of the critical path of the sorter is determined by the total number of stages which is computed based on the sum of the arithmetic progression as follows:

\begin{equation}
\label{eq:total_stages}
\Psi_{LD} = \sum_{\psi=1}^{1+\log_2 L} \psi=\frac{1}{2}(1+\log_2 L)(2+\log L) 
\end{equation}

From (\ref{eq:total_stages}), one can see the impact of the list size, $L$, on the latency of the sorter and consequently  the whole decoder. The pruned bitonic sorter suggested in \cite{balatsoukas-sorter} removes one stage out of $\Psi_{LD}$ stages, which is not significant in the case of large $L$, although the pruned network reduces the silicon area in hardware implementation. An efficient solution for a significant reduction in the latency is to employ list VA where the sorting is performed locally at each state. Therefore, the parameter $L$ in (\ref{eq:total_stages}) is divided by the number of states. It turns out the the order of the sorted metric in LVA is not needed unlike in the pruned bitonic sorter where the pruning is performed based on our prior knowledge about the order and the relations between adjacent metric before and after the tree extension. Hence, we can remove the last $log_2L$ stages in the last super-stage As a result, the total number of stages is:

\begin{equation}
\label{eq:total_stages_lva}
\Psi_{LVA} = 
\frac{1}{2}\left(1+\log_2 \frac{L}{2^{m}}\right)\left(2+\log \frac{L}{2^{m}}\right)-\log_2L 
\end{equation}

Thus, list VA results in a significant reduction in the latency of the decoding. For instance, for list decoding of PAC(256,128) with $m=6$ and $L=128$ which has 128 survivors at each decoding stage, the total number of sorting stages throughout decoding is $K  \Psi_{LD}=128\times 36=4608$. However, in decoding of the same code under list VA with $m=4$ and $L=128/2^4=8$ which has 32 survivors at each decoding stage, $K\Psi_{LVA}=128\times (10-3)=896$, which is 80\% smaller than its counterpart. Note that this reduction comes at the cost of a slight degradation in the FER performance. In a software implementation, the sorting algorithms such as Heapsort and Mergesort cannot perform better than $O(2L\log (2L))$ in terms of time complexity. By employing LVA, the time complexity reduces to  $O(2^m 2L/2^m\log (2L/2^m))=O(2L\log (2L/2^m))$.

\begin{figure}[t] 
    \centering
    \includegraphics[width=1\columnwidth]{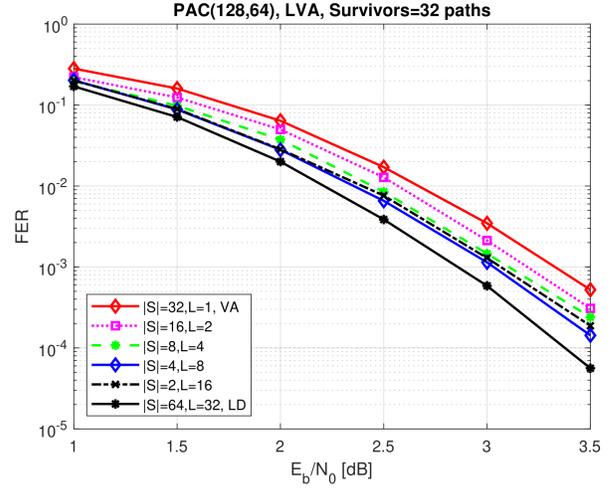}
    \caption{FER Comparison of LVA with various parameters while the total number of paths is 32.}
    \vspace{-5pt}
    \label{fig:FER1}
\end{figure}

\begin{figure}[t] 
    \centering
    \includegraphics[width=1\columnwidth]{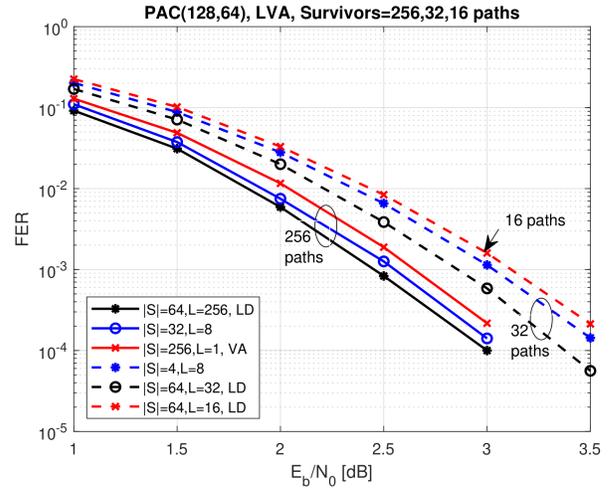}
    \caption{FER Comparison of LVA with various parameters while the total number of paths are 256, 32, and 16.} 
    \vspace{-5pt}
    \label{fig:FER2}
\end{figure}

\section{Numerical Results} \label{sec:results}
In this Section, the error correction performance of list Viterbi algorithm for PAC(128,64) on the trellis with different setups is illustrated 
and analyzed. The RM rate-profile \cite{rowshan-pac1} and the generator polynomials 0o3, 0o7, 0o17, 0o33, 0o73, and 0o133 ($m=1,\ldots, 6$), are used for convolutional transform (pre-coding) for the results shown in Fig. \ref{fig:FER1}  corresponding to a number of states  $|\mathcal{S}|= 2, 4, 8, 16, 32$ and 64 (for LD), respectively. 
Also, for the results shown in Fig. \ref{fig:FER2}, the generator polynomials 0o133, 0o73 and 0o733 are used for convolutional transform with a number of states $|\mathcal{S}|=64$ (for LD), 32 and 256, respectively. The codewords are modulated based on BPSK and transmitted over the AWGN channel. 
Fig. \ref{fig:FER1} compares the FER performance under LVA with various list sizes $L$, while the total number of survivor paths at each time step $t$ remains constant (32 survivors). As can be seen, the performance improves as $L$ increases. 
Fig. \ref{fig:FER2}  shows that as the total number of survivors increases, the gap between the performance of LD, VA and LVA decreases. This makes LVA a better candidate when employing a very large list size, given latency advantage shown in Section \ref{sec:sorting}. Conversely, when list size is small, the performance of LVA with list size $L$ is close to the performance of LD with list size $L/2$ as it is shown in Fig. \ref{fig:FER2} for LVA with $L=32$ and LD with $L=16$.

\section{Conclusion} 
In this paper, we investigate the implementation of the list Viterbi decoding  for PAC codes. We show that LVA can be considered a general decoding scheme, which can transition from list decoding to Viterbi algorithm decoding by changing the number of states and the local list size. The results show that as the local list size increases, the performance improves. This implies that in the local sorting of the paths, the probability of discarding the correct path is higher than the global sorting in list decoding. On the other hand, the local sorting has the advantage of a significantly lower latency than global sorting. Therefore, depending on the application, we can trade latency for performance, specially  when the list size is large.








\begin{thebibliography}{9}
\bibitem{arikan} E. Ar\i kan, ``Channel polarization: A method for constructing capacity-achieving codes for symmetric binary-input memoryless channels," {\em IEEE Trans. Inf. Theory,} vol. 55, no. 7, pp. 3051-3073, Jul. 2009.
\bibitem{tal} I. Tal and A. Vardy, ``List decoding of polar codes," {\em IEEE Int. Symp. on  Information Theory,} St. Petersburg, Russia, Jul. 2011, pp. 1–5.
\bibitem{arikan2} E. Ar\i kan, ``From sequential decoding to channel polarization and back again," arXiv preprint arXiv:1908.09594 (2019).
\bibitem{li2} B. Li, H. Zhang, J. Gu, ``On Pre-transformed Polar Codes," arXiv preprint arXiv:1912.06359 (2019).
\bibitem{rowshan-pac1} M. Rowshan, A. Burg and E. Viterbo, ``Polarization-adjusted Convolutional (PAC) Codes: Fano Decoding vs List Decoding," arXiv preprint  arXiv:2002.06805 (2020).
\bibitem{viterbi} A. Viterbi, ``Error bounds for convolutional codes and an asymptotically optimum decoding algorithm," in {\em  IEEE Transactions on Information Theory}, vol. 13, no. 2, pp. 260-269, April 1967.
\bibitem{forney} G.D. Forney, ``The Viterbi Algorithm," {\em Proc. of the IEEE}, Vol. 61, No. 3, pp. 268-278, Mar. 1973.
\bibitem{arikan_viterbi} E. Arıkan, H. Kim, G. Markarian, U. Ozgur and E. Poyraz, ``Performance of short polar codes under ML decoding," in {\em Proc. ICT-Mobile Summit Conf.}, Santander, Spain, 2009, pp. 1-6.
\bibitem{hashimoto} T. Hashimoto, ``A list-type reduced-constraint generalization of the Viterbi algorithm," in {\em IEEE Transactions on Information Theory}, vol. 33, no. 6, pp. 866-876, November 1987.
\bibitem{seshadri} N. Seshadri and C.-E. W. Sundberg, ``List Viterbi decoding algorithms with applications," in {\em IEEE Transactions on Communications}, vol. 42, no. 234, pp. 313-323, Feb-Apr 1994.
\bibitem{rowshan-stepped} M. Rowshan and E. Viterbo, ``Stepped List Decoding for Polar Codes," 2018 {\em IEEE 10th International Symposium on Turbo Codes \& Iterative Information Processing (ISTC)}, Hong Kong, Hong Kong, 2018, pp. 1-5.
\bibitem{yuan} B. Yuan and K. K. Parhi, ``Successive cancellation list polar decoder using log-likelihood ratios," {\em 2014 48th Asilomar Conference on Signals, Systems and Computers}, Pacific Grove, CA, 2014, pp. 548-552.
\bibitem{balatsoukas} A. Balatsoukas-Stimming, M. Bastani Parizi, and A. Burg, ``LLR-based successive cancellation list decoding of polar codes," {\em IEEE Trans. Signal Processing,} vol. 63, no. 19, pp. 5165-5179, Oct 2015.
\bibitem{belman} R. E. Bellman and S. E. Dreyfus, ``Applied Dynamic Programming", Princeton University Press, Princeton, NJ, 1962.


\bibitem{batcher} K. E. Batcher, ``Sorting networks and their applications," in {\em Proc. AFIPS Spring Joint Comput. Conf.}, vol. 32, 1968, pp. 307-314.
\bibitem{lin} J. Lin and Z. Yan, ``Efficient list decoder architecture for polar codes," in {\em Proc. IEEE Int. Symp. on Circuits and Systems (ISCAS)}, Jun. 2014, pp. 1022–1025.
\bibitem{balatsoukas-sorter} A. Balatsoukas-Stimming, M. Bastani Parizi and A. Burg, ``On metric sorting for successive cancellation list decoding of polar codes," {\em 2015 IEEE International Symposium on Circuits and Systems (ISCAS)}, Lisbon, 2015, pp. 1993-1996.
\end{thebibliography}
\end{document}